\documentclass[aps,prl,twocolumn,superscriptaddress,showpacs,draft]{revtex4}
\usepackage{graphicx}
\input{epsf}

\usepackage{amsmath}
\usepackage{amssymb}

\begin{document}

\title{Superinsulator as a phase of bi-particle localized states}

\author{ J. Lages }
\affiliation{\mbox{Institut UTINAM, UMR CNRS 6213, 
Universit\'e de Franche-Comt\'e, 25030 Besan\c{c}on, France}}
\author{D.L. Shepelyansky}
%\homepage[]{http://www.quantware.ups-tlse.fr}
\affiliation{\mbox{Laboratoire de Physique Th\'eorique du CNRS (IRSAMC), 
Universit\'e de Toulouse, UPS, F-31062 Toulouse, France}}

%\date{\today}
\date{November 30, 2010; updated December 4, 2010}

%\pacs{PACS numbers: 05.45.-a, 05.45.Ac, 05.45.Jn}
%\PACS{
%{05.45.-a}{Nonlinear dynamics and chaos}
%\and
%{05.45.Ac}{Low-dimensional chaos}
%{05.45.Jn}{High-dimensional chaos}
%}

\pacs{74.20.-z, 71.30.+h, 72.15.Rn}
\begin{abstract}
We propose a physical picture of superinsulator observed
recently in experiments with superconducting films
in a magnetic field. On the basis of previous numerical studies
we argue that a moderate attraction creates bi-particle localized
states at intermediate disorder strength when noninteracting electron
states are delocalized and metallic. Our present numerical study
show that such localized pairs are broken by a static electric field
which strength is above a certain threshold. 
We argue that such a breaking of localized pairs by a static
field is at the origin of superinsulator breaking
with a current jump observed experimentally
above a certain critical voltage.
\end{abstract}

\maketitle

\section{I. Introduction}

\begin{figure}
\centerline{\epsfxsize=7.9cm\epsffile{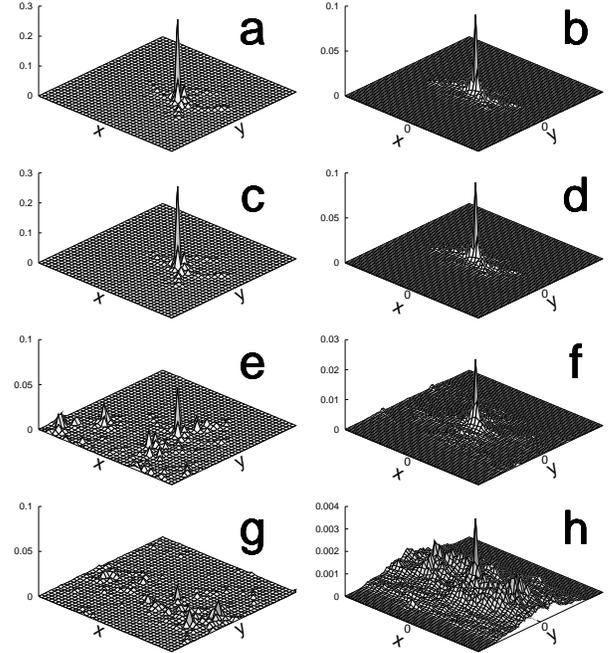}}
\vglue -0.2cm
\caption{Delocalization of two interacting particles
with attractive Hubbard interaction $U=-2V$ 
by a static electric force $F$.
The generalized Cooper problem is considered
on 2D lattice of size $L \times L = 40 \times 40$ 
at disorder strength $W=5V$, a static field
$F$ is directed along $y-$axis. Probability is shown for a lowest energy
eigenstate with a maximum probability $f(y)=\sum_x f(x,y)$
at $y=L/2$. Left panels show
one-particle probability $f(x,y)$ and right panels show 
interparticle distance probability $f_d(x,y)$. The static electric force,
directed along $y-$axis,  is
$F=0 (a,b)$, $F=0.003 V (c,d)$, $F=0.016 V (e,f)$,
$F=0.052 V (g,h)$.
} 
\label{fig1}
\end{figure}

An interplay of disorder, Anderson localization and superconductivity
attracts active experimental and theoretical interest
(see e.g. reviews \cite{goldman,gantmakher}
and a recent research article \cite{ioffe}).
A weak disorder does not significantly affect the superconducting phase
in agreement with the Anderson theorem \cite{anderson1959,degennes}.
However, a relatively strong disorder can lead to a nontrivial
situation when an attraction between electrons
creates bi-particle localized states (BLS phase) 
from noninteracting metallic delocalized electron states
\cite{lages2000,lages2001} (see Fig.~\ref{fig1}a,b). 
First signatures of this BLS phase
has been obtained in the frame of the generalized Cooper 
problem \cite{cooper} 
of two interacting particles above a frozen Fermi sea
in presence of disorder. Further extensive quantum Monte Carlo studies
of  two and three-dimensional (2D, 3D)
Hubbard model \cite{bhargavi2d,bhargavi3d}, 
with attraction, disorder and about hundred electrons,
confirmed the attraction induced picture of localized Cooper pairs
proposed  in \cite{lages2000,lages2001}. 
This picture is in a qualitative agreement with 2D disordered 
films experiments on superconductor-insulator transition (SIT) of 
Gantmakher {\it et al.} \cite{gantmakher1998}, which
clearly display a presence of localized pairs 
in the insulating phase 
appearing from the superconducting phase at
relatively large magnetic field 
(see also \cite{gantmakher} and Refs. therein).
At even larger magnetic fields these samples show
a metallic type behavior which is argued to 
correspond to underlying metallic noninteracting states
\cite{lages2001}.

Recent experiments also discovered that in 2D disordered films
the above insulating phase
is abruptly destroyed by a static {\it dc-}voltage
\cite{shahar2005,baturina2007,ovadyahu,baturinanature,shahar2009}.
It was argued that this unusual insulating phase 
is related to a certain collective state named
{\it superinsulator} \cite{baturinanature}. 
The physical origin of this state is 
under hot theoretical debates 
\cite{baturinanature,fistul,efetov,altshuler,efetov2}.
However, this physical problem
involves nontrivial interplay of interactions,
disorder and localization which
is not easy to handle by purely analytical methods.
Indeed, it is known that repulsive or attractive
interactions can produce delocalization of 
two interacting particles above Fermi level
when all one-particle states are exponentially localized due
to the Anderson localization 
(see e.g. \cite{dls1994,imry,pichard,jacquod,frahm,lagesmoriond}).
This two interacting particles (TIP) effect
leads to an effective 3D Anderson transition for
TIP excitations  at certain energy above the Fermi level 
in the case of Coulomb or  other long range interactions 
\cite{dls2000,lagestip}.
But at the same time  attractive interactions
create the BLS phase in the vicinity of Fermi level
even when noninteracting states are metallic and delocalized
\cite{lages2000,lages2001,bhargavi2d,bhargavi3d}.
Due to that for a better understanding of 
physics of superinsulator
it would be rather useful to study
numerically  
the effects of a static electric field
on localization-delocalization transition in presence of
interactions.  A certain progress has been reached
in studies of attractive Hubbard model with disorder
by quantum Monte Carlo methods since there is no
sign problem in such a case and numerical simulations
can be done with a large number of electrons
(see e.g. \cite{bhargavi2d,bhargavi3d,trivedi2010}
and Refs. therein). However, this approach is not 
easy to adapt to a case of finite static field.

In this work we study effects of static field
in the frame of the generalized Cooper problem
using the approach of two interacting particles
with an attractive Hubbard interaction $U$ \cite{lages2000,lages2001}.
Qualitative description of a 
physical picture of BLS phase and its destruction 
by a static electric field $F$ is given in Section II.
In Section III we describe our model of the generalized Cooper problem
in presence of {\it dc-}field, the results of numerical simulations
are presented in Section IV,
discussion of physical results and comparison with
experiments are given in Section V.

\section{II. Physical picture of superinsulator}

The results obtained in \cite{lages2000,lages2001,bhargavi2d,bhargavi3d}
give the following qualitative physical picture of the BLS phase,
which is at the origin of superinsulator as it is argued below:

{\it i)} For a moderate disorder, which e.g. by a factor two or less
smaller than the critical value for the Anderson transition for 3D case,
noninteracting electron states are still delocalized and metallic.
A similar situation appears for finite size 2D samples
where one-particle localization length is larger than the sample size.

{\it ii)} However, an attractive Hubbard interaction creates
singlet spin pairs of two electrons which total mass is twice larger than
electron mass, hence, an effective hopping matrix element of such a pair
becomes twice smaller or, equivalently, effective strength of disorder 
becomes twice larger that leads to localization of pairs and BLS phase.
Such localized pairs have a certain localization size $\ell$
and a coupling energy $\Delta$. 
It is important to stress that the Bogoliubov - de Gennes meanfield
approach is not able to capture this BLS phase 
\cite{lages2000,lages2001,bhargavi2d,bhargavi3d}.
The BLS phase is an insulator existing
at temperature $T < \Delta$. 
In a certain sense attraction between two electrons
creates an effective well which captures localized electron pairs.
However, excitations 
with energy $\Delta E > E_D \approx \Delta$
are delocalized. Indeed, even if all one-particle states
are localized,
the excited states above the Fermi level
become delocalized by
interactions between electrons
as it is shown in \cite{jacquod,lagesmoriond,dls2000,lagestip}.
Moreover, 
the energy excitations of BLS phase are delocalized above certain 
energy $E_D \approx \Delta$ since noninteracting states
are metallic.
According to the Fermi-Dirac statistical distribution
an excitation probability 
to high energy drops exponentially and therefore
for $T < \Delta$ the resistivity $R_{xx}$
is characterized by the Arrhenius law $R_{xx} \propto \exp(T_0/T)$
with $T_0 \approx E_D \approx \Delta$.
\begin{figure}
\centerline{\epsfxsize=6.9cm\epsffile{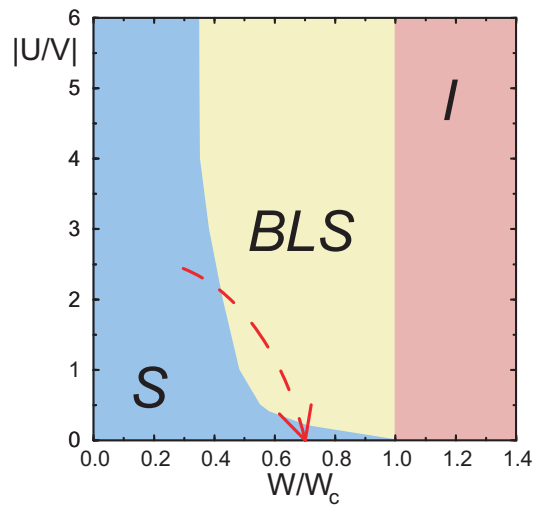}}
\vglue -0.0cm
\caption{(Color online) Phase diagram in the 3D Anderson model
with disorder strength $W$ and Hubbard attraction $U$:
superconductor ($S$), phase of localized pairs ($BLS$),
Anderson insulator ($I$); $W_c \approx 16.5V$ is the critical disorder 
for noninteracting particles at half filling.
The data, taken from \cite{lages2000},
are obtained in the frame of the generalized Cooper problem
with TIP.
The dashed curve with arrow schematically shows variation of
effective $U$ and $W$ values with the increase of magnetic field
corresponding to the experiment \cite{gantmakher1998}
(see Section II for more details). This phase diagram is 
shown for zero temperature $T$.
} 
\label{fig2}
\end{figure}

{\it iii)} In real 2D superconducting films, studied experimentally
(see e.g. \cite{gantmakher1998,gantmakher}), an increase of magnetic field $B$
up to a few Tesla effectively decreases attraction inside
Cooper pairs due to breaking of time reversal and also effectively
increases an effective strength of the disorder
since magnetic field increases a return probability and 
scattering on impurities.
As a result superconductivity disappears, electron pairs become
localized, but at even stronger magnetic field
attraction between electrons is completely eliminated and one obtains
a metal of noninteracting electrons. 
This effective change of attraction  $U$
and disorder strength $W$ with an increase of magnetic field $B$
is schematically shown in Fig.~\ref{fig2} by a dashed curve.
In this picture, 
resistivity $R_{xx}$ initially grows with increase of $B$
but above a certain value it starts to decrease
with $B$ giving a peak in $R_{xx}$,
which is a characteristic feature of experiments
(see e.g. Fig.~1 in \cite{gantmakher1998}).

\begin{figure}
\centerline{\epsfxsize=6.9cm\epsffile{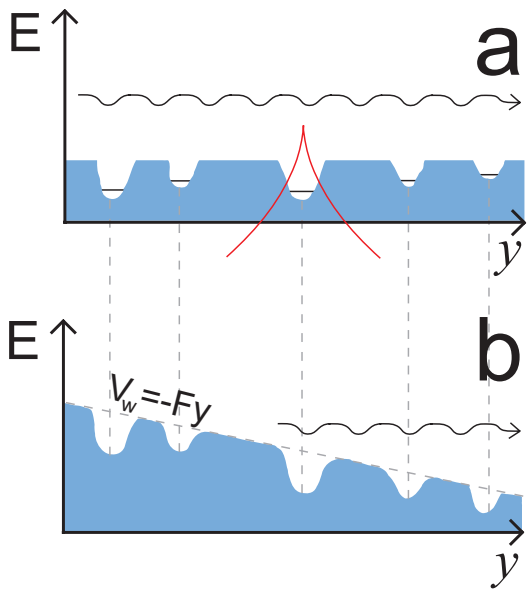}}
\vglue -0.0cm
\caption{(Color online) Pictorial image of the 
energy spectrum of BLS pairs localized by
a Hubbard attraction and located at various
positions along $y-$direction of the lattice at static force $F=0$ (a);
delocalization of BLS pairs by a 
washboard potential $V_w=-Fy$ of
static field at $F>F_c$ (b).
} 
\label{fig3}
\end{figure}

{\it iv)} Thus, the BLS phase and its energy excited states
can be viewed as a sequence of wells with localized pairs
and continuum of delocalized states as it is schematically shown
in Fig.~\ref{fig3}a. A static electric field with force $F$
creates an energy slope leading to a tilted washboard potential
as it is schematically shown in Fig.~\ref{fig3}b.
Above a certain critical force $F>F_c$
the localized states inside a well 
(localized pairs) become coupled to continuum
states of delocalized electrons 
that creates an avalanche of delocalized
electrons. For $F>F_c$ all electrons become delocalized
that produces a sharp increase (jump) of electron current
which is a characteristic feature of 
superinsulator experiments 
\cite{shahar2005,baturina2007,ovadyahu,baturinanature,shahar2009}.
The critical field $F_c$ above which the superinsulator is destroyed
can be estimated by taking into account that the coupling energy
of localized pairs $\Delta$ should be comparable with 
the energy change in a static field $F_c$ on the pair size $\ell$
that gives
\begin{equation}
\label{eq1}
F_c \approx \Delta/ \ell  \; , \;\;V_c = F_cL
\end{equation}
The physical meaning of this relation is rather direct:
a strong field breaks BLS pairs and creates a  charge current.
A critical voltage $V_c$,
at which a jump of current takes place in an experiment, 
is proportional to the sample size $L$.

In next Sections we present  numerical simulations
of the generalized Cooper problem in a tilted potential
which justifies this physical picture.

\section{III. Model description}
For our numerical studied of delocalization of BLS pairs by a static
electric field $F$ we use 2D Anderson model with
disorder strength $W$ and Hubbard attraction $U$
between two particles. Following \cite{lages2001}, with the same notations,
we use the one-particle Hamiltonian:
\begin{equation}
\label{eq2}
H_{1}=\sum_{\mathbf n} (E_{\mathbf n} +{\mathbf {F}\cdot\mathbf{n}})\left\arrowvert
{\mathbf n}\left\rangle\right\langle{\mathbf n}\right\arrowvert
+ V\sum_{\langle\mathbf{n,n'}\rangle} 
\left\arrowvert {\mathbf n}
\left\rangle\right\langle{\mathbf n'}\right\arrowvert 
\end{equation}
where $\mathbf n$ and $\mathbf n'$ are index vectors on the 
two-dimensional square lattice
with periodic boundary conditions in $x-$direction, 
and zero boundary conditions in $y-$direction,
$V$ is the nearest neighbor hopping
term and the random on-site energies $E_{\mathbf n}$ are homogeneously
distributed in the energy interval 
$\left[-\frac{W}{2},\frac{W}{2}\right]$. We choose the direction of 
a static electric force ${\mathbf F}$ along $y-$axis.
We consider a square lattice with linear size $L$ up to $L=40$.
At such sizes the eigenstates at a half filling $\nu =1/2 $
are practically delocalized over the whole lattice
for $W<7V$ and $F=0$ so that such finite samples can be considered
to be metallic (see more details in \cite{lages2001}).  
We consider
the particles in the singlet state with zero total spin so that the
spatial wavefunction is symmetric with respect to particle permutation
(interaction is absent in the triplet state). 

To take into account the effects of Hubbard interaction
we write the TIP Hamiltonian in the basis of 
noninteracting eigenstates at $F=0$:
\begin{eqnarray}
\label{eq3}
&( E_{m_1}& +  E_{m_2})\chi_{m_1, m_2}  + 
\displaystyle\sum_{m'} (F_{m_1,m'} \chi_{m', m_2} + \nonumber\\ 
&F_{m_2,m'}&\chi_{m_1, m'} )   
+U \displaystyle\sum_{{m^{'}_1}, {m^{'}_2}} Q_{m_1, m_2, {m^{'}_1}, {m^{'}_2}}
 \chi_{{m^{'}_1}, {m^{'}_2}} \nonumber \\ 
  & =&  E\chi_{m_{1}, m_{2}}. 
\end{eqnarray}
Here $E_{m}$ are the one-particle eigenenergies corresponding to the 
one-particle eigenstates $\arrowvert\phi_m\rangle$ and 
$\chi_{m_1,m_2}$ are the components of the TIP eigenstate in the
non-interacting eigenbasis $\arrowvert\phi_{m_1}, \phi_{m_2}\rangle$
at $F=0$. The matrix elements $F_{m_1,m'}$ describe
the static force transitions between one-particle eigenstates
$\arrowvert\phi_{m_1}, \phi_{m_2}\rangle$.  
The matrix elements $UQ_{m_1, m_2, {m^{'}_1}, {m^{'}_2}}$ give
the interaction induced transitions between non-interactive 
eigenstates $\arrowvert\phi_{m_1}, \phi_{m_2}\rangle$ and 
$\arrowvert \phi_{m^{'}_1}, \phi_{m^{'}_2}\rangle$. These matrix
elements are obtained by rewriting the Hubbard interaction 
in the non-interactive 
 eigenbasis of model (\ref{eq2}) at $F=0$.
In the analogy with the original Cooper problem \cite{cooper}
the summation in (\ref{eq3}) is done over the states above
the Fermi level with eigenenergies $E_{m^{'}_{1,2}}>E_F$ with 
$m^{'}_{1,2}>0$. The Fermi energy $E_F\approx 0$ is determined 
by a fixed filling factor $\nu =1/2 $. To keep the similarity
with the Cooper problem we restrict the summation on $m^{'}_{1,2}$
by the condition $1< m^{'}_1+m^{'}_2\leq M$. In this way the 
cut-off with $M$ unperturbed orbitals introduces an effective
phonon energy 
$E_{ph} =\hbar\omega_D \approx 3.75 V M/L^2= 3.75 V/\alpha$ where $L$
is the linear system size. When varying $L$ we keep $\alpha=L^2/M$
fixed so that the phonon energy is independent of system
size. All the data in this work are obtained with $\alpha=15$
but we also checked that the results are not sensitive to
the change of $\alpha$. We note that the Hamiltonian (\ref{eq3})
exactly describes the noninteracting problem. 

To analyze the effects of static force on localized BLS pairs
we solved numerically the Schr\"odinger equation
(\ref{eq3}). After that we rewrite the obtained 
eigenstates
in the original lattice basis with the help of the relation
between lattice basis and one-particle eigenstates
$\arrowvert {\mathbf n}\rangle= \sum_m R_{{\mathbf n},m} 
\arrowvert \phi_m\rangle$. As a result of this procedure
we obtain the two-particle probability distribution 
$f_2({\mathbf {n_1,n_2}})$ from which we extract the one-particle 
probability 
$f({\mathbf n})=\sum_{\mathbf n_2} f_2({\mathbf {n_1,n_2}})$ 
and the probability of interparticle distance 
$f_d({\mathbf r})=\sum_{\mathbf n_2} f_2({\mathbf {r+n_2,n_2}})$
with ${\mathbf {r=n_1-n_2}}$. The localization properties are characterized
by the one-particle inverse participation ratio (IPR)
$\xi = \sum_{{\mathbf n}} f({\mathbf n})/\sum_{{\mathbf n}} f^2({\mathbf n})$.

\begin{figure}
\centerline{\epsfxsize=7.9cm\epsffile{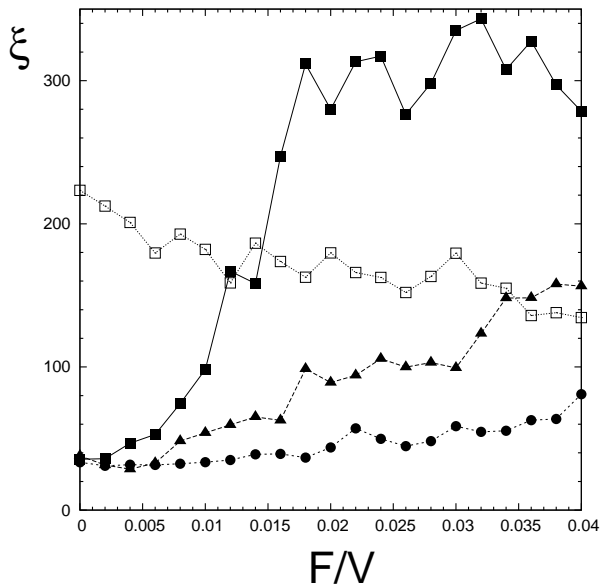}}
\vglue -0.2cm
\caption{Dependence of the average
IPR $\xi$ of lowest energy states
on the static force $F$; average is done over
$N_l=15$ lowest energy
states with a maximum of $f(x,y)$
inside a stripe $0.4 L \leq y \leq 0.6L$ in a
middle of the lattice at $y=L/2$ for $N_D=30$ disorder realizations;
%(only states with a TIP overlap
%$f_d(0,0) > 5/L(2J-1)$ are considered);
the lattice size is
$L=20$ ($\bullet$), $L=30$ ($\blacktriangle$), $L=40$ ($\blacksquare$)
at $U=-2V$; the values of $\xi$ for noninteracting case are shown 
by open  symbols ($\square$) for $L=40, U=0$. Here $W=5V$.
} 
\label{fig4}
\end{figure}

While in \cite{lages2001} only the ground state properties 
of given disorder realization have been studied here
we also investigate the properties of excited states
with the TIP energy $\Delta E$ counted
from the Fermi level of noninteracting particles:
$\Delta E = E - 2E_F$, where $E$ is the eigenenergy
of (\ref{eq3}).
We also consider only eigenstates with a maximum
of one-particle probability inside the space range
$-L/4 \leq y \leq L/4$ to avoid finite size effects in $y-$direction.
In addition, we  analyze only those eigenstates where the
overlap probability of TIP to be on the same site is
relatively large $f_d(0,0) > 5/L(2L-1)$.
In this way the states with strongly separated particles are eliminated.
Such an approach approximately corresponds to a finite particle density.
We use usually $N_D=30$ disorder realizations
for statistical average. 

Below we present numerical results for $U=-2V$, $W=5V$ and $L \leq 40$
at various values of $F$. The detailed studies presented 
in \cite{lages2001} ensure that 
at $F=0$ these conditions are located well
inside the BLS phase when the  noninteracting states are delocalized
(see Fig.~1b in \cite{lages2001})
while the ground state in presence of Hubbard attraction is well
localized (see Fig.~1e,f in \cite{lages2001} and Fig.~\ref{fig1}a,b here). 
We checked that the behavior in $F$ remains similar at
other values of parameters, e.g. $W=3V$.

\section{IV. Numerical results}

The delocalization of BLS pairs by a static electric force $F$
is illustrated in Fig.~\ref{fig1} for
one specific disorder realization. Here the disorder strength
is relatively weak so that at $U=0$ noninteracting eigenstates
taken at half filling $\nu=1/2$ and $E_F \approx 0$
are delocalized over the whole lattice of size $L=40$
(see Fig.~1b in \cite{lages2001}). At $F=0$ 
a moderate  Hubbard attraction $U=-2V$
creates localized pairs with a certain coupling energy $\Delta$
(Fig.~\ref{fig1}a,b). 
This localization remains robust
against a weak static field (Fig.~\ref{fig1}c,d)
but at larger fields the localization is destroyed
by a static force and particles become delocalized 
over the whole lattice  (Fig.~\ref{fig1}e,f,g,h;
to avoid boundary effects at finite $F$
we select states in the middle of the lattice at $y=L/2$).
The probability to have two particles close to each other
also drops drastically for $F>F_c$
clearly demonstration pair breaking.

\begin{figure}
\centerline{\epsfxsize=7.9cm\epsffile{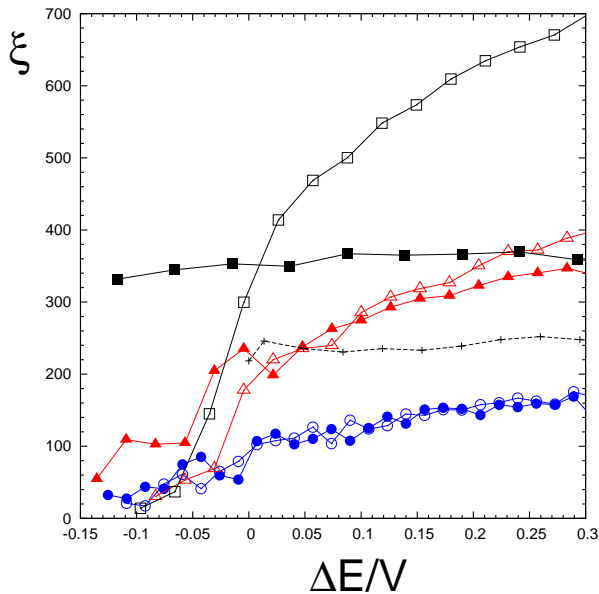}}
\vglue -0.2cm
\caption{(Color online) Average IPR $\xi$, for states peaked 
inside a stripe $0.4 L \leq y \leq 0.6 L$ ,
as a function of the coupling energy $\Delta E=E-2E_F$ 
for $U=-2V,W=5V$ and lattice size 
$L=20$ (blue circles), $L=30$ (red triangles), $L=40$ (black squares), 
at field $F=0$ (open symbols) and $F=0.036V$ (full symbols). 
The same averaged IPR $\xi$ 
for $U=0,W=5V,F=0$ and $L=40$ is shown by ($+$) symbols.
} 
\label{fig5}
\end{figure}

The dependence of average IPR $\xi$ of lowest energy states
on the static force $F$ is shown in Fig.~\ref{fig4}.
At small $F < F_c$ the values of $\xi$ are size independent 
being much smaller compared to 
the case of noninteracting particles.
This shows that a Hubbard attraction gives localization of pairs
at low energy.
For $F>F_c$ IPR
starts to grow with the system size $L$
demonstrating breaking of pairs and particle delocalization 
over the whole lattice. According to the data of Fig.~\ref{fig4}
we have $F_c \approx 0.015V$ at given $U=-2V$ and $W=5V$.
The data in Fig.~\ref{fig5} give the coupling energy
$\Delta \approx 0.1 V$.
Hence, this $F_c$ value is in a good agreement with a simple estimate
(\ref{eq1}) with a numerical factor
$A=F_c \ell /\Delta \approx 1$ corresponding to
$\ell(F=0) \approx \sqrt{\xi(F=0)} \approx 6$ and $\Delta=0.1V$.

\begin{figure}
\centerline{\epsfxsize=7.9cm\epsffile{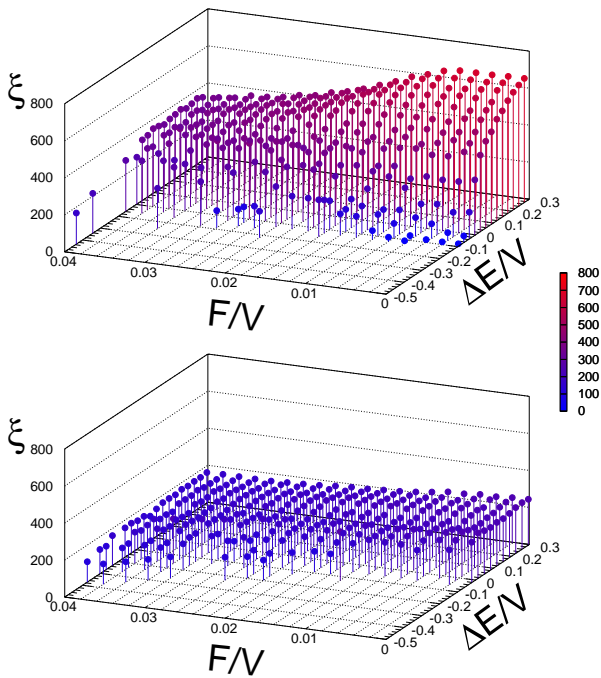}}
\vglue -0.2cm
\caption{(Color online)  Average IPR $\xi$, for states peaked 
inside a stripe $0.4 L \leq y \leq 0.6 L$ ,
as a function of the coupling energy $\Delta E=E-2E_F$ 
and static force $F$
for $U=-2V,W=5V$ (top panel) and  $U=0,W=5V$ (bottom panel);
here $L=40$.
} 
\label{fig6}
\end{figure}

The dependence of $\xi$ on coupling energy 
$\Delta E = E - 2E_F$, for states in the middle of 
the lattice at $y \approx L/2$, is shown in Fig.~\ref{fig5}.
According to this data the coupling energy is
$\Delta \approx 0.1V$. This value is by a factor 2 smaller
than the one found in \cite{lages2001}
since only one lowest state 
for a given disorder realization was
taken in \cite{lages2001}
while here we average over few lowest states
and also allow a relatively weak overlap $f_d$ between 
TIP states.
The states with 
$-\Delta \leq \Delta E <0$ are well localized at $F=0$
since its IPR $\xi$ is independent of lattice size $L$.
In contrast, for $F=0.036V > F_c$
the IPR $\xi$ grows with the system size $L$
showing that in this regime the states are delocalized.

At energies $\Delta E >0$ the IPR grows with energy
and becomes comparable with the system size $L^2$;
also its is not  sensitive to $F$.
This is in agreement with the fact that noninteracting 
states are delocalized.
Also interaction for excited states gives 
an additional TIP delocalization.  
For large values of $L=40$ the static force  
gives a  certain restriction of 
eigenstates spreading along the force direction
due to the TIP energy conservation
that gives a decrease of IPR value comparing to 
the case $F=0$.

\begin{figure}
\centerline{\epsfxsize=7.9cm\epsffile{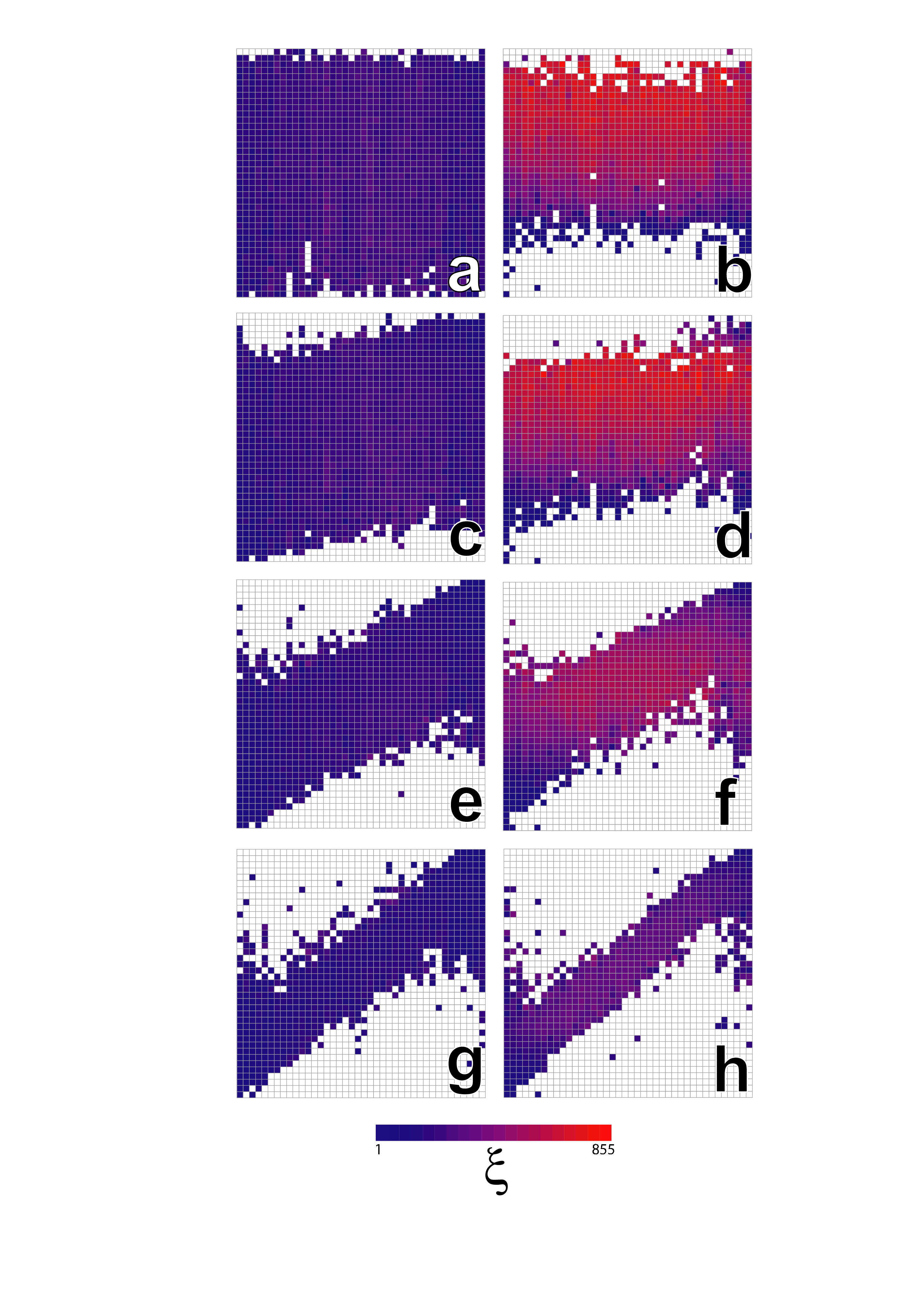}}
\vglue -0.2cm
\caption{(Color online) 
The IPR $\xi$ 
 for $W=5V$ at $U=0$ (left column) and $U=-2V$ (right column), 
at different values of the static force 
$F=0$ (a,b), $F=0.004V$ (c,d), $F=0.016V$ (e,f), $F=0.04V$ (g,h).
Each panel has $40\times40$ cells, 
the vertical direction corresponds to the coupling energy 
$\Delta E=E-2E_F$, the horizontal one to $1 \leq y \leq 40$. 
The bottom row corresponds to the lowest  coupling energy, 
and the upper row corresponds to the highest coupling energy
within the energy intervals 
at $U=0$ being
$(0,0.54V)$ (a), $(-0.05V,0.59V)$ (c),
$(-0.25V,0.78V)$ (e), $(-0.69V,1.18V)$ (g)
and at $U=-2V$ being
$(-0.2V, 0.39V)$ (b),
$(-0.23V,0.57V)$ (d), $(-0.39V,0.75V)$ (f), $(-0.77V, 1.17V)$ (h).
The cell  color gives the average $\xi$ inside the cell
(with $\Delta E$ being in the corresponding energy range defined by the row, 
and  the maximum of  probability distribution  $f(y)$ along $y$, 
being located at $y$ position defined by the column). 
For each panel $\sim 30000$ states  are used with
$N_d=30$ disorder realizations. 
These states are selected in such a way that the probability 
of two particles  located at the same site is greater 
than $5/L(2L-1)$; here $L=40$. 
} 
\label{fig7}
\end{figure}

A more detailed dependence of $\xi$ on coupling energy 
$\Delta E$ and static force $F$ is given in Fig.~\ref{fig6}.
For $U=0$ we have $\xi \approx 200 $ which is practically independent
of $\Delta E$ and $F$ while for $U=-2V$
we have very small $\xi \sim 10$ for $F=0$ 
and large $\xi \sim 300$ for large $F$ at 
$-0.1 <\Delta E/V <0$. For energies $\Delta E > 0$
the states are delocalized at all $F$.
These data also confirm the picture of 
field induced destruction of the BLS phase and delocalization.

Dependence of $\xi$ on $y$ and coupling energy $\Delta E $
is shown in Fig.~\ref{fig7} for different values of $F$
with and without interaction.
For $U=0$ we have $\xi$ practically independent of $y$ and $\Delta E$
in agreement with previous data of Figs.~\ref{fig4},\ref{fig5},\ref{fig6}.
In contrast, in presence of attraction the states in the middle of 
the lattice in $y$ have small $\xi$ (localized) 
at lowest energies for small $F$ (panels b,d)
and have large $\xi$ (delocalized) for large F (panels f,h).
However, at the ends of the lattice in $y$ direction
the values of $\xi$ are less sensitive to $F$
due to boundary effects. Indeed, the static field forces 
particles to stay close to the boundary at $y-$ends of the lattice
and hence the field induced delocalization is not well visible
in this region. Due to that reason we use the states
in the middle of the lattice to detect field induced delocalization
in a clear way in Figs.~\ref{fig4},\ref{fig5},\ref{fig6}.

\section{V. Discussion}
\begin{figure}
\centerline{\epsfxsize=7.9cm\epsffile{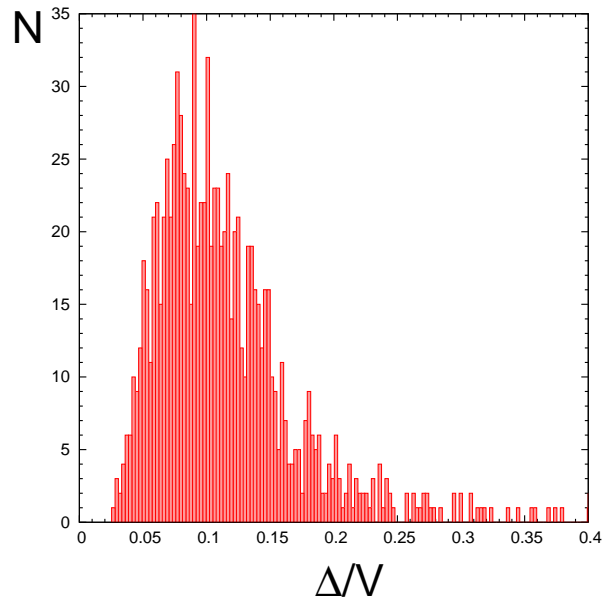}}
\vglue -0.2cm
\caption{(Color online) Distribution histogram of pair coupling energy
$\Delta= 2E_F - E_g$ obtained from $N_D=1000$
disorder realizations at $U=-2V$, $W=5V$, $F=0$, $L=40$;
here $N$ gives a number of realizations found in a given cell
of $\Delta/V$, $E_g$ is the ground state energy
of a given realization obtained numerically from (\ref{eq3}). 
} 
\label{fig8}
\end{figure}

The obtained numerical results confirm the physical picture
of superinsulator destruction by a static field
described in Section II: the BLS pairs, localized
by attraction inside noninteracting metallic phase,
in presence of static field
start to be coupled with higher energy delocalized states
and above certain threshold $F>F_c$ (\ref{eq1})
all pairs become delocalized. This creates
an avalanche of delocalized electrons which gives
an enormous increase of current in agreement with
experimental observations.
Of course, our numerical data detect delocalization of only
one pair in a given disorder realization.
However, the distribution of values of pair coupling 
energy $\Delta=2E_F-E_g$
is strongly peaked near its average value (see Fig.~\ref{fig8})
so that a large fraction of localized pairs becomes
delocalized approximately at the same static field
that gives a sharp current growth for $V>V_c =F_c L$
(here $E_g$ is a ground state energy for a given disorder
realization at $F=0$).

At that point we would like to note that 
even if our attraction is formally relatively strong
(e.g. $|U|=2V \sim V$)  
it effectively gives a rather weak
coupling energy
of BLS pairs $\Delta \approx 0.1 V$ (see Fig.~\ref{fig8}).
There are a few physical reasons behind this.
At first, a simple physical estimate gives a consistent value 
$\Delta \sim |U|/\xi \sim 0.05V \ll V$ where 
we use numerically found value of IPR
$\xi \approx \ell^2 \approx 40$ from Fig.~\ref{fig4}.
Thus the numerical value of $\Delta$ is by a factor 100 smaller 
than the energy band width of the noninteracting 2D problem
$E_b \approx 8V+W \approx 14V$.
Second, in (\ref{eq3}) the summation over one-particle orbitals
is done only inside a Debye energy interval
$E_{ph}=\hbar \omega_D \approx 3.75 V /\alpha \approx 0.25 V \ll E_b$
and due to that the effective attraction is additionally decreased 
giving a relatively small coupling energy.
Also in a good metallic phase (e.g. $W \leq 2V$) the interaction
(e.g. even $|U|=2V$) produces a quite weak effect on
noninteracting delocalized states
according to numerical results presented
in \cite{lages2000,lages2001}
and according to usual theoretical estimates
for interaction matrix elements 
$U_s \sim U/g \ll E_b$, which are inversely
proportional to a sample conductance $g \gg 1$
(see e.g. \cite{imry,jacquod} and Refs. therein). 
Due to these reasons we think that the claim
expressed in \cite{ioffe}, that the BLS pairs
appear only as a result of nonphysically strong attraction,
is not justified since our BLS pairs have
rather small coupling energy $\Delta \ll E_b$ and have
rather large size $\xi \gg 1$.
In addition, the numerical data presented in 
\cite{lages2000} (see also Fig.~\ref{fig2})
show that the BLS pairs appear also at smaller
values of attraction. Thus, even if, with the aim 
to have well localized pairs inside
a system size available for numerical 
simulations, we fixed an attraction at a relatively strong value,
our numerical data show that such a choice is still
in the regime of weakly coupled pairs of relatively larger size
that corresponds to the experimental regime.
We note that even larger $|U|/V$ values are typically used 
in quantum Monte Carlo simulations (see e.g. \cite{trivedi2010}).

\begin{figure}
\centerline{\epsfxsize=7.9cm\epsffile{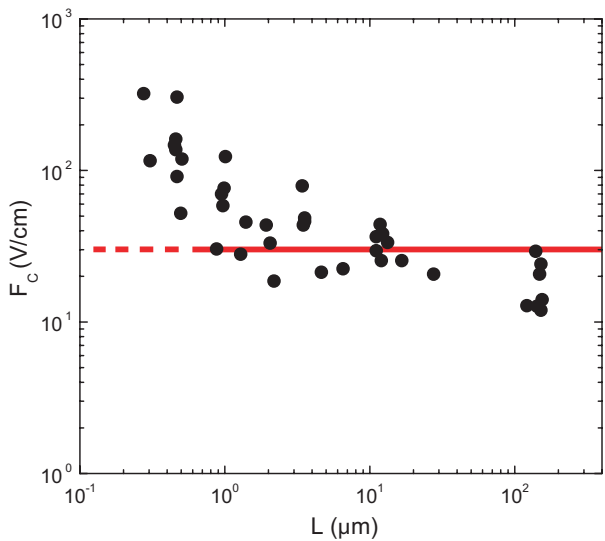}}
\vglue -0.2cm
\caption{(Color online) Critical field $F_c$ for
superinsulator destruction as a function of sample size
$L$ from experiments \cite{ovadyahu} (points from Fig.5 there).
The theory  (\ref{eq1}) is shown by red horizontal line.
} 
\label{fig9}
\end{figure}

The main result of this studies is given by 
Eq.~(\ref{eq1}) which determines the critical voltage $V_c$
of superinsulator destruction via the sample size $L$,
BLS coupling energy $\Delta$ and pair size $\ell$.
According to (\ref{eq1}) the critical voltage $V_c$
is proportional to the sample size and hence,
$F_c$ is independent of $L$. This is 
in a good agreement with the experimental data obtained
in \cite{ovadyahu} (see Fig.~\ref{fig9} with experimental points
from Fig.5 in \cite{ovadyahu}).
Indeed, in the range $0.5 \mu m < L <150 \mu m$
the value of $F_c$ shows significant fluctuations
but in average remains constant. The average value is
$F_c \approx 30 V/cm$, and since the typical
value of pair coupling energy is
$\Delta \approx T_0 \approx 3 - 15 K$, we find the 
size of localized pairs to be $\ell = T_0/F_c = 100 - 500 n m$.
The theory (\ref{eq1}) is valid for $L > \ell$ where indeed $F_c$
is independent of $L$, a part of fluctuations. 
However, for $L < \ell \sim 0.5 \mu m$ one enters into another regime
where the sample size becomes comparable with the pair size
that can lead to an increase of $F_c$ seen experimentally.
It is clear that as soon as the superinsulator phase 
is destroyed at $F>F_c$ a further decrease of $F$ below
$F_c$ places localized pairs in other new locations
where  due to fluctuations of disorder
one gets a comparable but somewhat different new value of $F_c$.
This leads to a hysteresis behavior observed experimentally.

It is also interesting to note that the relation (\ref{eq1})
allows to determine the dependence of $\ell$ on magnetic field.
Indeed, in experiments \cite{shahar2005,baturina2007,baturinanature}
the values of $L$ is known, and also $\Delta \approx T_0$ and  $V_c$ 
are experimentally
known as a function of magnetic field $B$
that allows to determine the dependence $\ell(B)$
from the relation (\ref{eq1}).  

\begin{figure}
\centerline{\epsfxsize=7.9cm\epsffile{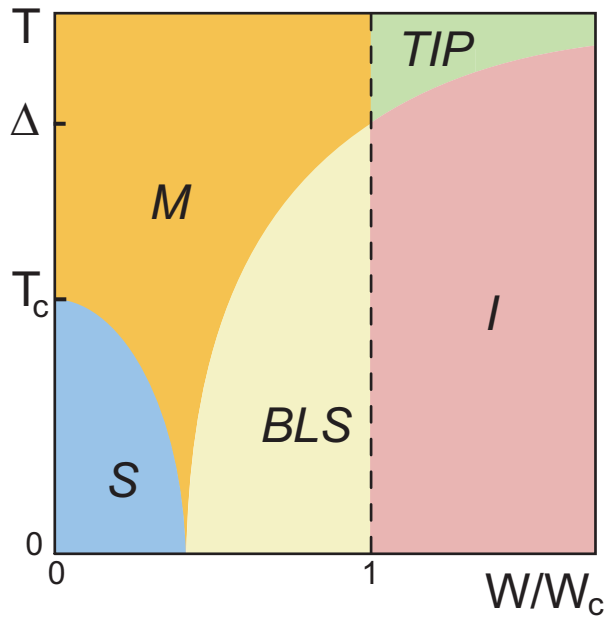}}
\vglue -0.2cm
\caption{(Color online) Schematic phase diagram 
in the temperature-disorder plane
$(T,W)$ for quasi-2D or 3D Anderson model 
at fixed Hubbard attraction $U=-2V$ and 
fixed filling factor with the Fermi energy $E_F$:
the vertical dashed line shows the Anderson transition
at $W=W_c \approx 16.5V$ for noninteracting particles,
$S$ is the superconducting phase,
$BLS$ is the insulator phase of localized BLS pairs,
$M$ is essentially the noninteracting
metallic phase, $TIP$ is the metallic phase
of delocalized TIP pairs. Here we assume $T \ll E_F$,
static field is zero,
temperature linear axis is shown in arbitrary units.
} 
\label{fig10}
\end{figure}

On the basis of presented results we can draw a global
phase diagram in the temperature-disorder plane
shown in Fig.~\ref{fig10}
for a fixed attraction strength 
(e.g. at $U=-2V$), zero static field and fixed Fermi energy $E_F$.
At small disorder and temperature we have the superconducting 
phase $S$ which is followed by a transition to 
metallic phase $M$ at large temperature or to
the localized BLS phase at low temperature.
At a larger disorder $W>W_c$, 
but still small temperatures,
the BLS phase
enters in the insulating regime of
noninteracting Anderson insulator $I$.
In this regime with $W> W_c$ 
but temperature above a certain 
threshold 
$T > T_2 \sim |U| (1/\xi+1/\xi_1)$, the TIP pairs
become delocalized by interactions
with emergence of metallic phase
of TIP delocalized pairs,
as it is discussed in \cite{dls1994,imry,jacquod}.
Here $\xi_1$ is a noninteracting one-particle IPR
which gives a dominant contribution
for a disorder $W>W_c$ which is not
very close to the critical point $W=W_c$,
the term $1/\xi$ with IPR of BLS is included
to have interpolation between two phases. 

In conclusion, we presented the BLS based physical
picture for a destruction of superinsulator
by a finite static field observed experimentally in 
\cite{shahar2005,ovadyahu,baturina2007,baturinanature,shahar2009}.
This picture is rather different from other theoretical
explanations discussed in 
\cite{baturinanature,fistul,efetov,altshuler,efetov2}.
The main new element of our theory is the existence
of localized pairs created by attraction inside noninteracting
metallic phase which is absent in the above theoretical
proposals. In our theory the BLS phase is the basis of superinsulator
and since the noninterating states are metallic this phase is
sharply broken by a finite static field 
which breaks electron pairs, localized by attraction, and
lets them propagate like noninteracting particles in a metal. 
Our analysis considers a breaking of only two interacting
particles. In a real system with many pairs
a breaking of a few pairs can create a strong avalanche
and breaking of other pairs so that a critical
static field can be determined by breaking of mostly weakly 
coupled pairs with a smaller critical fields,
compared to average field values found here.
This physical picture is rather different 
from the superinsulator picture discussed in 
\cite{baturinanature}. However, we keep the term
superinsulator which in our opinion nicely
describes impressive experiments
\cite{shahar2005,baturina2007,baturinanature,shahar2009}
with superconducting films, which become
insulating in magnetic fields with
abrupt emergence of  current above a certain
critical ${\it dc-}$voltage.

This research is supported in part by the ANR PNANO project
NANOTERRA. We thank V.F.~Gantmakher for useful remarks and
M.V~Feigel'man for constructive critical comments.

\end{document}